\documentclass[10pt,conference]{IEEEtran}
\usepackage{enumerate}
\usepackage{amsthm}
\usepackage{amsmath}
\usepackage{amscd}
\usepackage{dsfont}
\usepackage{amssymb}
\usepackage{latexsym}
\usepackage{array}
\usepackage{tabularx}
\usepackage{tikz}
\usepackage{graphicx,subfigure}
\usepackage{bbm}
\usepackage[numbers, square, comma, compress]{natbib}

\newtheorem{deft}{Definition}

\theoremstyle{definition}
\newtheorem{rem}{Remark}
\newtheoremstyle{dotless}{}{}{\itshape}{}{\bfseries}{}{ }{}
\theoremstyle{dotless}

\newcommand{\Z}{\mathsf{Z}}

\newcommand{\R}{\mathbb{R}}

\newcommand{\X}{\mathsf{X}}
\newcommand{\Y}{\mathsf{Y}}
\newcommand{\K}{\mathsf{K}}

\newcommand {\aplt} {\ {\raise-.5ex\hbox{$\buildrel<\over{\mbox{\scriptsize $\sim$}}$}}\ }
\providecommand{\abs}[1]{\ensuremath{\left\lvert #1 \right\rvert}}

\DeclareMathOperator*{\av}{av}

\DeclareMathOperator*{\Mod}{\,mod}

\makeatletter \g@addto@macro\@floatboxreset\centering \makeatother

\begin{document}

\title{Secret key generation from Gaussian sources \\using lattice hashing}

\author{\IEEEauthorblockN{Cong Ling}
\IEEEauthorblockA{Department of Electrical \\ and Electronic Engineering \\
Imperial College London, U.K.\\
cling@ieee.org} \and
\IEEEauthorblockN{Laura Luzzi}
\IEEEauthorblockA{Laboratoire ETIS\\
 CNRS - ENSEA - UCP \\
Cergy-Pontoise, France \\
laura.luzzi@ensea.fr} \and
\IEEEauthorblockN{Matthieu R. Bloch}
\IEEEauthorblockA{School of ECE\\
Georgia Institute of Technology, \\
Atlanta, GA, U.S.A.\\
matthieu.bloch@ece.gatech.edu}
}

\maketitle

\begin{abstract}
We propose a simple yet complete lattice-based scheme for secret key generation from Gaussian sources in the presence of an eavesdropper, and show that it achieves strong secret key rates up to $1/2$ nat from the optimal in the case of ``degraded" source models. The novel ingredient of our scheme is a lattice-hashing technique, based on the notions of flatness factor and channel intrinsic randomness. The proposed scheme does not require dithering.
\end{abstract}

\section{Introduction}

Secret key generation at the physical layer was first investigated by Maurer \cite{Maurer93a} and Ahlswede and Csisz\'ar \cite{AC_IT93}, who showed that correlated observations of noisy phenomena could be used to distill secret keys by exchanging information over a public channel. Most existing secret key generation schemes rely heavily on the assumption of discrete random sources over finite or countable alphabets. In order to apply these techniques to wireless communications, it is necessary to extend the key generation framework to the case of continuous sources, such as Gaussian sources \cite{WaOh10,Nitinawarat_Narayan12}. In \cite{Nitinawarat_Narayan12}, the authors study a multi-terminal scenario for secret key generation in the special case where the eavesdropper only has access to the public channel. Beside providing a characterization of the optimal strong secret key rate, they show that this optimal rate can be achieved using lattice codes (for information reconciliation only).

In this paper, we consider secret key generation between two terminals, Alice and Bob, who observe correlated Gaussian sequences $\X^n$ and $\Y^n$, in the presence of an eavesdropper, Eve, who also obtains a correlated sequence $\Z^n$. For the sake of simplicity, we suppose that a single round of unidirectional public communication takes place in order to establish the key.
Our main contribution is to show that, in the case of a ``degraded" source model, a secret key rate up to half a nat from the optimal can be achieved by a complete lattice-coding scheme considerably different from and much simpler than \cite{Nitinawarat_Narayan12}.

Typically, secret key generation is composed of two distinct procedures: \emph{information reconciliation}, in which public messages are exchanged to ensure that Alice and Bob can construct the same data sequence with vanishing error probability, and \emph{privacy amplification} to extract from this shared sequence a secret key which is statistically independent from Eve's observation and from the public messages.
\subsubsection*{Privacy amplification and randomness extraction} Our privacy amplification strategy is based on the concept of \emph{channel intrinsic randomness},
or the maximum bit rate that can be extracted from a channel output independently of its input  \cite{Bloch_Intrinsic_Randomness, MuKoMu03, Hayashi11}).
We begin by considering a simplified scenario in which Bob and Alice share the same variable $\X^n$. In this case, the amount of randomness which can be extracted from $\X^n$ independently of $\Z^n$ is precisely the maximum available secret key rate.
We propose a \emph{lattice-hashing} technique to extract the randomness, by reducing the source modulo a suitable lattice. Although our main objective in this paper is to solve the problem of privacy amplification, our lattice-hashing technique is also an intriguing result in its own right, which could have other applications.
\subsubsection*{The flatness factor} Furthermore, we provide a quantitative characterization of the class of lattices which are good for randomness extraction, which is based on a computable parameter, the \emph{flatness factor}, as a function of the eavesdropper's noise variance. The concept of flatness factor relates to the properties of Gaussian measures on lattices, and was first introduced in \cite{BelfioreITW11} in the context of physical-layer network coding. In \cite{LLBS_12}, two of the authors also showed the relevance of the flatness factor for secrecy and introduced the notion of \emph{secrecy-good lattices} for the wiretap channel. In this paper, we show that this class of lattices is also suitable for secret key generation.
\subsubsection*{Information reconciliation and Wyner-Ziv coding} Our strategy for information reconciliation follows the outline of \cite{WaOh10,Nitinawarat_Narayan12}: first, the source $\X^n$ is vector quantized; then, a public message is generated in the manner of Wyner-Ziv coding, so that Bob can decode the quantized variable using the sequence $\Y^n$ as side information.  The existence of good nested lattices for Wyner-Ziv coding has been established in \cite{Zamir02} (see also \cite{LCLX06,LingCOM12}). We show that this construction is compatible with the secrecy-goodness property to conclude our existence proof.
\subsubsection*{Organization}
This paper is organized as follows. In Section \ref{flatness_factor_section} we provide basic definitions about lattices and introduce the flatness factor, which allows to define the notion of secrecy-good lattices. In Section \ref{extractor_section}, we focus on the extraction of channel intrinsic randomness over Gaussian channels using lattice hashing. In Section IV, we introduce the Gaussian source model and describe our lattice-based secret key generation scheme. The existence of sequences of nested lattices satisfying the required conditions is shown in Section \ref{existence_section}.

\section{Lattices and Flatness Factor} \label{flatness_factor_section}

In this section, we introduce the mathematical tools we use to describe and analyze our
proposed scheme.

An $n$-dimensional {lattice} $\Lambda$ in the Euclidean space
$\mathbb{R}^{n}$ is the discrete set defined by
\begin{equation*}
\Lambda=\mathcal{L}\left( \mathbf{B}\right) =\left\{ \mathbf{Bx}\text{ : }\mathbf{x\in }\text{ }%
\mathbb{Z}^{n}\right\}
\end{equation*}%
where the columns of the basis matrix $\mathbf{B=}\left[ \mathbf{b}_{1}\cdots \mathbf{b}_{n}\right]
$ are linearly independent.

A measurable set~$\mathcal{R}(\Lambda)\subset \mathbb{R}^n$ is called a fundamental region of the
lattice~$\Lambda$ if the disjoint union~$\cup_{\lambda \in \Lambda} (\mathcal{R}(\Lambda)+\lambda) = \R^n$.
Examples of fundamental regions
include the fundamental parallelepiped $\mathcal{P}(\Lambda)$ and the Voronoi region $\mathcal{V}(\Lambda)$. All the fundamental regions have equal
volume~$V(\Lambda)$.

For a vector $\mathbf{x}$, the mod~$\mathcal{R}(\Lambda)$ operation is defined by
$\mathbf{x}\mapsto \bar{\mathbf{x}}$ where $\mathbf{\bar{x}}$ is the unique element of
$\mathcal{R}(\Lambda)$ such that $\bar{\mathbf{x}} - \mathbf{x} \in \Lambda$.
For simplicity, we write this operation as
\begin{equation}\label{eq:mod}
\bar{\mathbf{x}}= \mathbf{x} \Mod \mathcal{R}(\Lambda).
\end{equation}
For any $\mathbf{x} \in \R^n$, the nearest-neighbor quantizer associated with $\Lambda$ is given by $Q_{\Lambda}(\mathbf{x})=\arg\min_{\lambda \in\Lambda}\|\lambda-\mathbf{x}\|$. Note that $\mathbf{x} \Mod \mathcal{V}(\Lambda)=\mathbf{x}-Q_{\Lambda}(\mathbf{x})$.

Suppose that $\X^n$ is an $n$-dimensional i.i.d. Gaussian random variable of variance $\sigma^2$ with distribution
\begin{equation*}
 f_{\sigma}(\mathbf{x})=\frac{1}{(\sqrt{2\pi}\sigma)^n}e^{- \frac{\|\mathbf{x}\|^2}{2\sigma^2}},
\end{equation*}
for $\mathbf{x} \in\R^n$. We consider the $\Lambda$-periodic function
\begin{equation}\label{Guass-function-lattice}
  f_{\sigma,\Lambda}(\mathbf{x})=\frac{1}{(\sqrt{2\pi}\sigma)^n}
\sum_{\lambda \in \Lambda} e^{-
    \frac{\|\mathbf{x}+\lambda\|^2}{2\sigma^2}},
\end{equation}
for all $\mathbf{x} \in\R^n$. Observe that $f_{\sigma,\Lambda}$ restricted to $\mathcal{R}(\Lambda)$ is the probability density of $\bar{\mathbf{x}}$.

\begin{deft} [Flatness factor \cite{LLBS_12}]
For a lattice~$\Lambda$ and for a parameter~$\sigma$, the flatness factor
is defined by:
\begin{equation*}
\epsilon_{\Lambda}(\sigma)  \triangleq \max_{\mathbf{x} \in
\mathcal{R}(\Lambda)}\abs{
V(\Lambda)f_{\sigma,\Lambda}(\mathbf{x})-1}
\end{equation*}
\end{deft}
In other words, $\epsilon_{\Lambda}(\sigma)$ characterizes the $L^{\infty}$ distance of $f_{\sigma,\Lambda}(\mathbf{x})$ to the uniform distribution $U_{\mathcal{R}(\Lambda)}$  over~$\mathcal{R}(\Lambda)$.

The flatness factor can be computed from the theta series $\Theta_{\Lambda}$ of the lattice, using the identity \cite{LLBS_12}
\begin{equation} \label{flatness_factor_expression}
\epsilon_{\Lambda}(\sigma) =  \left(\frac{\gamma_{\Lambda}(\sigma)}{{2\pi}}\right)^{\frac{n}{2}}{
\Theta_{\Lambda}\left({\frac{1}{2\pi\sigma^2}}\right)}-1,
\end{equation}
$\gamma_{\Lambda}(\sigma) = \frac{
V(\Lambda)^{\frac{2}{n}}}{\sigma^2}$ is the volume-to-noise ratio (VNR).

\begin{rem} \label{monotonicity}
We have shown in \cite{LLBS_12} that $\epsilon_{\Lambda}$ is a monotonically decreasing function, i.e., for $\sigma < \sigma'$, we have $\epsilon_{\Lambda}(\sigma') \leq  \epsilon_{\Lambda}(\sigma)$.
\end{rem}

The notion of secrecy-goodness
characterizes lattice sequences whose flatness factors vanish exponentially fast
as $n \to \infty$.

\begin{deft}[Secrecy-good lattices \cite{LLBS_12}] \label{secrecy_goodness}
A sequence of lattices $\Lambda^{(n)}$
is \emph{secrecy-good} if
$\epsilon_{\Lambda^{(n)}}(\sigma)  = e^{-\Omega(n)}$
for all fixed $\gamma_{\Lambda^{(n)}}(\sigma)<2\pi$.
\end{deft}

In \cite{LLBS_12} we have proven the existence of sequences of secrecy-good lattices as long as
\begin{equation} \label{secrecy_condition}
\gamma_{\Lambda}(\sigma)<2\pi.
\end{equation}

\begin{rem} \label{concentration}
In fact, we can show a concentration result: $\forall \eta>0$ there exists a mod-$p$ lattice ensemble such that lattice sequences from this ensemble are secrecy-good with probability greater than $1-\eta$ (see \cite[Appendix III]{LLBS_12}).
\end{rem}

\section{Lattice Hashing for Gaussian Sources} \label{extractor_section}
Consider now a source model for secret key generation with public discussion, in the presence of an eavesdropper.
For simplicity, we first assume that Alice and Bob observe the same i.i.d. Gaussian
random variable $\X^n=\Y^n$ of variance $\sigma_x^2$ per dimension.
Eve observes a correlated i.i.d. random variable $\Z^n$.
We assume that
$\X^n$ and $\Z^n$ are jointly Gaussian,
according to the following model
\begin{equation}
\X^n=\Z^n+\mathsf{W}^n,
\label{model_Section3}
\end{equation}
where $\mathsf{W}^n$ is an i.i.d. zero-mean Gaussian random vector of variance $\sigma^2$ per dimension. We suppose that $\mathsf{W}^n$ and $\Z^n$ are independent.

Our aim is to extract from $\X^n$
a random number
that is almost uniform on $\mathcal{R}(\Lambda)$ and almost independent of $\Z^n$.
To do this, we apply the mod $\mathcal{R}(\Lambda)$ operation in Eq.~(\ref{eq:mod}). Then, the conditional
density of $\bar{\X}^n=\X^n \Mod \mathcal{R}(\Lambda)$ given $\Z^n$ is
\begin{align*}
& p_{\bar{\X}^n|\Z^n}(\bar{\mathbf{x}}|\mathbf{z}) =
\sum_{\mathbf{x}: \; \bar{\mathbf{x}}=\mathbf{x} \Mod \mathcal{R}(\Lambda)} p_{\X^n|\Z^n}(\mathbf{x}|\mathbf{z}) \\
&= \sum_{\mathbf{{x}}\in \bar{\mathbf{x}}+ \Lambda} p_{\X^n|\Z^n}(\mathbf{x}|\mathbf{z}) = \sum_{\lambda \in \Lambda} \frac{1}{(\sqrt{2\pi}\sigma)^n} e^{-
    \frac{\|\bar{\mathbf{x}}+\lambda-\mathbf{z}\|^2}{2\sigma^2}} \\
    &=f_{\sigma,\Lambda}(\bar{\mathbf{x}}-\mathbf{z}) \mathds{1}_{\mathcal{R}(\Lambda)}(\bar{\mathbf{x}}).
\end{align*}
From the definition of the flatness factor, it then follows that
\begin{equation}
\forall \mathbf{z} \in \R^n, \; \forall \bar{\mathbf{x}} \in \mathcal{R}(\Lambda), \quad \abs{p_{\bar{\X}^n|\Z^n}(\bar{\mathbf{x}}|\mathbf{z})-\frac{1}{V(\Lambda)}}\leq \frac{\epsilon_{\Lambda}(\sigma)}{V(\Lambda)}.
\label{bound_z}
\end{equation}

With a similar reasoning, we also find
$$p_{\bar{\X}^n}(\bar{\mathbf{x}})=f_{\sigma_x,\Lambda}(\bar{\mathbf{x}}) \mathds{1}_{\mathcal{R}(\Lambda)}(\bar{\mathbf{x}})$$
and again  by definition of the flatness factor, we find
\begin{align}
\forall \bar{\mathbf{x}} \in \mathcal{R}(\Lambda), \quad \abs{p_{\bar{\X}^n}(\bar{\mathbf{x}})-\frac{1}{V(\Lambda)}}\leq \frac{\epsilon_{\Lambda}(\sigma_x)}{V(\Lambda)}. \label{bound_xbar}
\end{align}
So, if the flatness factor is small,
$\bar{\X}^n$ is almost uniformly distributed over $\mathcal{R}(\Lambda)$, and also almost independent of $\Z^n$.

It is worth mentioning that unlike other works which use dithering or the high-resolution assumption \cite{Zamir02}, we obtain uniformity and independence from the flatness factor.

One can now bound the mutual information
{\allowdisplaybreaks
\begin{align*}
&\mathbb{I}(\bar{\X}^n;\Z^n) =
\int_{\R^n} \int_{\mathcal{R}(\Lambda)}{p_{\bar{\X}^n\Z^n}(\bar{\mathbf{x}},\mathbf{z}) \log{\frac{p_{\bar{\X}^n|\Z^n}(\bar{\mathbf{x}}|\mathbf{z})}{p_{\bar{\X}^n}(\bar{\mathbf{x}})}}}d\bar{\mathbf{x}}d\mathbf{z} \nonumber \\
&\leq  \int_{\R^n} \int_{\mathcal{R}(\Lambda)}{p_{\bar{\X}^n\Z^n}(\bar{\mathbf{x}},\mathbf{z}) \log{\frac{1+\epsilon_{\Lambda}(\sigma)}{1-\epsilon_{\Lambda}(\sigma_x)}}}d\bar{\mathbf{x}}d\mathbf{z} \nonumber \\
&= \log{(1+\epsilon_{\Lambda}(\sigma))} - \log{(1-\epsilon_{\Lambda}(\sigma_x))} \leq \epsilon_{\Lambda}(\sigma) + 2\epsilon_{\Lambda}(\sigma_x)
\end{align*}
}%
if $\epsilon_{\Lambda}(\sigma_x)\leq \frac{1}{2}$.
Since
$\sigma_x \geq \sigma$ by Eq.~ (\ref{model_Section3}), and recalling Remark \ref{monotonicity}, we have $\mathbb{I}(\bar{\X}^n;\Z^n)\leq 3 \epsilon_{\Lambda}(\sigma)$. The sufficient condition in Eq.~(\ref{secrecy_condition}) for the existence of secrecy-good lattices ensures that $\mathbb{I}(\bar{\X}^n;\Z^n)$
vanishes exponentially if
$\gamma_{\Lambda}(\sigma) <2\pi$.

Observe that depending on the choice of $\Lambda$, the rate of extracted randomness can be arbitrarily large.
\begin{rem} The asymptotic differential entropy rate of $\bar{\X}^n$ is
\begin{align*}
r=\liminf_{n \to \infty}\frac{1}{n} h(\bar{\X}^n) &\geq \liminf_{n \to \infty}\frac{1}{n}[\log{V(\Lambda)} - \log(1+\epsilon_{\Lambda}(\sigma_x))]
\end{align*}
Taking a sequence of secrecy-good lattices such that
$\gamma_{\Lambda^{(n)}}(\sigma) \to 2\pi$ as $n \to \infty$, which is compatible with the condition (\ref{secrecy_condition}), we can obtain the asymptotic rate
$r = \log(\sqrt{2\pi} \sigma)$, which is only $\frac{1}{2}$ nat from the asymptotic differential entropy rate of the Gaussian
noise $\mathsf{W}^n$ (i.e., $\log(\sqrt{2\pi e} \sigma)$). 
\end{rem}

Note that neither nearest-neighbor quantization nor dither is used in our lattice-hashing scheme, and we only need to implement the mod $\mathcal{R}(\Lambda)$ operation,
which can be performed in polynomial time for many fundamental regions $\mathcal{R}(\Lambda)$. In particular, we can choose the fundamental
parallelepiped. Moreover, Remark \ref{concentration} implies that if the lattice $\Lambda$ is chosen randomly in a mod-$p$ ensemble, it is secrecy-good with high probability, so that one can obtain explicit schemes.

\section{Secret Key Agreement} \label{secret_key_section}
From the above discussion, it seems that one can get an arbitrarily high rate of the secret key, since
$\bar{\X}^n$ is continuous.
However, this is fictitious, because
$\X^n \neq \Y^n$
in practice. This requires Alice and Bob to agree on the key over a public channel, which will lead to a finite key rate.

\begin{figure}[htb]
\begin{footnotesize}
\begin{tikzpicture}[
nodetype1/.style={
	rectangle,
	rounded corners,
	minimum width=10mm,
	minimum height=7mm,
	dashed,
	draw=black,
	text centered
},
nodetype2/.style={
	rectangle,
	rounded corners,
	minimum width=16mm,
	minimum height=7mm,
	text width=16mm,
	text centered,
	draw=black
},
tip2/.style={-latex,shorten >=0.4mm}
]
\matrix[row sep=0.6cm, column sep=0.7cm, ampersand replacement=\&]{
\node (Alice) {\textsc{Alice}}; \& \& \& \node (Bob) {\textsc{Bob}};\\
\node (key_gen1)  [nodetype2] {\textsc{key \\ generation}};  \& \node (quantizer) [nodetype2] {\textsc{quantizer}}; \& \node (source) [draw, nodetype1] {$p_{\X\Y\Z}$}; \& \node (key_gen2) [nodetype2] {\textsc{decoder}}; \\
\& \& \node (Eve) {\textsc{Eve}}; \& \\
};
\draw[->] (key_gen1) edge[tip2] node [right] {$\K$} (Alice);
\draw[->] (key_gen2) edge[tip2] node [right] {$\hat{\K}$} (Bob);
\draw[->] (source) edge[tip2]  node [above] (X) {$\X^n$} (quantizer);
\draw[->] (quantizer) edge[tip2]  node [above=-0.1cm] (XQ) {$\X^n_{Q}$} (key_gen1);
\draw[->] (source) edge[tip2]  node [above] (Y) {$\Y^n$} (key_gen2);
\draw[->] (source) edge[tip2]  node [right] (Z) {$\Z^n$} (Eve);
\draw (Eve.south) node[below=0.4cm] (dot) {};
\draw[-,>=latex] (dot.center) -| node [anchor=south,right,pos=0.9] {$\mathsf{S}$} (key_gen1);
\draw[->,>=latex] (dot.center) -| node [anchor=south, right,pos=0.9] {$\mathsf{S}$} (key_gen2);
\draw[>=latex] (dot.center) edge[tip2]  node [anchor=south, right] {$\mathsf{S}$} (Eve);
\draw (dot.south) node[below] {public channel (noiseless)};
\end{tikzpicture}
\end{footnotesize}
\caption{Secret key generation in the presence of an eavesdropper with communication over a public channel.}
\label{Figure_SecretKey}
\end{figure}
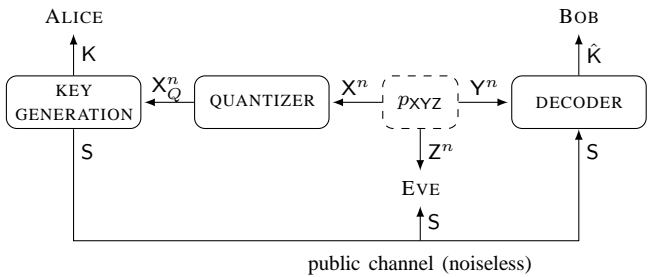

We consider an i.i.d. memoryless Gaussian source
$p_{\X\Y\Z}$
whose components are jointly Gaussian with zero mean. The distribution is fully described by the variances $\sigma_x^2$, $\sigma_y^2$, $\sigma_z^2$ and the correlation coefficients $\rho_{xy}$, $\rho_{xz}$, $\rho_{yz}$. We can write~\cite[Eq.~(6)]{WaOh10}:
\begin{equation}\label{XYZ-model}
\begin{split}
\X^n &= \rho_{xy}\frac{\sigma_x}{\sigma_y} \Y^n + \mathsf{W}_1^n, \\
\X^n &= \rho_{xz}\frac{\sigma_x}{\sigma_z} \Z^n + \mathsf{W}_2^n,
\end{split}
\end{equation}
where $\mathsf{W}_1^n$
and $\mathsf{W}_2^n$
are i.i.d. zero-mean Gaussian noise vectors of variances
\begin{equation}
\sigma_1^2=\sigma_x^2(1-\rho_{xy}^2),\quad \sigma_2^2=\sigma_x^2(1-\rho_{xz}^2),
\label{variances}
\end{equation}
respectively. Further,
$\mathsf{W}_1^n$ is independent of
$\Y^n$, and $\mathsf{W}_2^n$
is independent of
$\Z^n$.

The results of the previous section allow to extract from $\X^n$ a random variable $\bar{\X}^n$ that is almost statistically independent of $\Z^n$, with $\sigma^2$ replaced by $\sigma_2^2$. Note that the coefficient $\rho_{xz}\sigma_x/\sigma_z$ does not affect the argument based on the flatness factor. However, not all the extracted randomness can be exploited to generate the key, because Bob has to reconstruct
$\X^n$ with side information $\Y^n$, which requires Wyner-Ziv coding. Also, in secret key generation, we are not concerned with the standard rate-distortion function, but with the error probability of the key.

We assume that only one round of one-way public communication (from Alice to Bob) takes place. More precisely, Alice computes a public message $\mathsf{S}$ and a secret key $\K$ from her observation $\X^n$; she then transmits $\mathsf{S}$ over the public channel (see Figure \ref{Figure_SecretKey}). From this message and his own observation $\Y^n$, Bob reconstructs a key $\hat{\K}$.

Let $\mathcal{K}_n$ and $\mathcal{S}_n$ be the sets of secret keys and public messages respectively. A \emph{secret key rate - public rate pair} $(R_K,R_P)$ is achievable if there exists a sequence of protocols with
$$ \liminf_{n \to \infty} \frac{1}{n} \log \abs{\mathcal{K}_n} \geq R_K, \quad \limsup_{n \to \infty} \frac{1}{n}\log \abs{\mathcal{S}_n} \leq R_P, $$
such that the following properties hold:
\begin{align*}
& \lim_{n \to \infty} \log \abs{\mathcal{K}_n} - \mathbb{H}(\K) = 0 \quad &\text{(uniformity)} \\
&\lim_{n \to \infty} \mathbb{P} \left\{ \K \neq \hat{\K} \right\}=0 \quad &\text{(reliability)} \\
& \lim_{n \to \infty} \mathbb{I}(\K;\mathsf{S},\Z^n)=0\quad & \text{(strong secrecy)}.
\end{align*}
To define our key generation scheme, we use the lattice partition chain $\Lambda_1/\Lambda_2/\Lambda_3$, where
\begin{itemize}
  \item $\Lambda_1$ is quantization-good, which serves as the ``source-code" component of Wyner-Ziv coding;
  \item $\Lambda_2$ is AWGN-good, which serves as the ``channel-code" component in Wyner-Ziv coding;
  \item $\Lambda_3$ is secrecy-good with respect to $\sigma_2$, which serves as the extractor of randomness.
\end{itemize}
The existence of such a chain of lattices will be shown in Section \ref{existence_section}.

We suppose that the lattices are scaled so that their volumes $V_1$,$V_2$,$V_3$ satisfy
$$\abs{\Lambda_2/\Lambda_3}=\frac{V_3}{V_2}=e^{nR_K}, \quad \abs{\Lambda_1/\Lambda_2}=\frac{V_2}{V_1}=e^{nR_P}.$$
The procedure of secret key generation is described as follows:
\begin{itemize}
  \item Alice quantizes
 $\X^n$
  to
  $\X^n_Q=Q_{\Lambda_1}(\X^n) \in \Lambda_1$. She then computes $$\mathsf{S}=\X^n_Q \Mod \mathcal{V}(\Lambda_2),$$ which belongs to a set of coset leaders of $\Lambda_1/\Lambda_2$ in $\mathcal{V}(\Lambda_2)$,
  and transmits its index to Bob. Furthermore, Alice computes the key
  $$\K=Q_{\Lambda_2}(\X^n_Q) \Mod \mathcal{R}(\Lambda_3),$$
  which belongs to a set of coset leaders of $\Lambda_2/\Lambda_3$ in $\mathcal{R}(\Lambda_3)$.
    Note that
    \begin{equation}
\X^n=\mathsf{E}_Q^n+\mathsf{S}+\K + \lambda_3
\label{X}
\end{equation}
for some $\lambda_3 \in \Lambda_3$, where $\mathsf{E}^n_Q=\X^n-\X^n_Q \in \mathcal{V}(\Lambda_1)$ is the quantization error.
  \item Bob receives $\mathsf{S}$
  and reconstructs
  $$\hat{\X}_Q^n=\mathsf{S}+Q_{\Lambda_2}\left(\rho_{xy}\frac{\sigma_x}{\sigma_y}\Y^n-\mathsf{S}\right).$$
 He then computes his version of the key
 $$\hat{\K}=Q_{\Lambda_2}(\hat{\X}_Q^n) \Mod \mathcal{R}(\Lambda_3).$$
\end{itemize}

Note that
$\K$ and $\mathsf{S}$ are functions of $\bar{\X}^n=\X^n \Mod \mathcal{R}(\Lambda_3)$: generalizing~\cite[Eq.~(35)]{Nazer11}, we have
\begin{align*}
\K&=Q_{\Lambda_2}(Q_{\Lambda_1}(\X^n)) \Mod \mathcal{R}(\Lambda_3)=\notag\\
&=Q_{\Lambda_2}(Q_{\Lambda_1}(\X^n) \Mod \mathcal{R}(\Lambda_3)) \Mod \mathcal{R}(\Lambda_3)= \notag\\
&=(Q_{\Lambda_2}(Q_{\Lambda_1}(\bar{\X}^n) \Mod \mathcal{R}(\Lambda_3))) \Mod \mathcal{R}(\Lambda_3)=f(\bar{\X}^n);
\end{align*}
moreover, it is not hard to see
that
$\X^n \Mod \mathcal{V}(\Lambda_2)=\bar{\X}^n \Mod \mathcal{V}(\Lambda_2)$, and thus
\begin{align*}
&\mathsf{S}=
Q_{\Lambda_1}(\X^n \Mod \mathcal{V}(\Lambda_2)) \Mod \mathcal{V}(\Lambda_2)=\\
&=Q_{\Lambda_1}(\bar{\X}^n) \Mod \mathcal{V}(\Lambda_2)=g(\bar{\X}^n).
\end{align*}

\subsubsection*{Uniformity}
Using the results of the previous section, we can show that
$\K$
is almost uniformly distributed on $\Lambda_2 / \Lambda_3$: from Eq.~(\ref{X}) we have $\bar{\X}^n=\X^n \Mod \mathcal{R}(\Lambda_3)=(\mathsf{E}_Q^n+\mathsf{S}+\K) \Mod \mathcal{R}(\Lambda_3)$, and $\forall k \in \Lambda_2 \cap \mathcal{R}(\Lambda_3)$,
$$p_{\K}(k)=\sum_{s \in \mathcal{S}_n}\int_{\mathcal{V}(k+s)} p_{\bar{\X}^n}(\bar{\mathbf{x}}) d\bar{\mathbf{x}}.$$
where $\mathcal{V}(k+s)=(\mathcal{V}(\Lambda_1)+k+s) \Mod \mathcal{R}(\Lambda_3)$. Using the bound in Eq.~(\ref{bound_xbar}), we find that $\forall k \in \Lambda_2 \cap \mathcal{R}(\Lambda_3)$,
\begin{equation}
\abs{p_{\K}(k) - \frac{V_2}{V_3}} =\abs{p_{\K}(k) - \frac{1}{e^{nR_K}}} \leq \frac{\epsilon_{\Lambda_3}(\sigma_x)}{e^{nR_K}}.
\label{bound_K}
\end{equation}
Consequently, the entropy of the key is lower bounded by
\begin{align*}
&\mathbb{H}(\K) \geq \sum_{k \in \Lambda_2 \cap \mathcal{R}(\Lambda_3)} p_{\K}(k) \log\left(\frac{e^{nR_K}}{1+\epsilon_{\Lambda_3}(\sigma_x)}\right)=\\
&=nR_K -\log(1+\epsilon_{\Lambda_3}(\sigma_x)) \geq nR_K -\epsilon_{\Lambda_3}(\sigma_x).
\end{align*}

\subsubsection*{Strong secrecy}
We recall the following bound from \cite{Csiszar96}:
$$\mathbb{I}(\K;\mathsf{S},\Z^n) \leq d_{\av} \log\frac{\abs{\mathcal{K}_n}}{d_{\av}},$$
where
$$d_{\av}=\sum_{k \in \mathcal{K}_n} p_{\K}(k) \mathbb{V}(p_{\mathsf{S}\Z^n|\K=k},p_{\mathsf{S}\Z^n}),$$
and $\mathbb{V}$ denotes the variational distance.

Observe also that $(\mathsf{S},\K)\to \bar{\X}^n \to \Z^n$ is a Markov chain.
Therefore we have
{\allowdisplaybreaks
\begin{align*}
&p_{\mathsf{S}\Z^n|\K=k}(s,\mathbf{z}|k)=\frac{p_{\mathsf{S}\Z^n\K}(s,\mathbf{z},k)}{p_{\K}(k)}=\\
&=\frac{1}{p_{\K}(k)}\int_{\mathcal{R}(\Lambda_3)} p_{\mathsf{S}\Z^n\K|\bar{\X}^n}(s,\mathbf{z},k|\bar{\mathbf{x}})p_{\bar{\X}^n}(\bar{\mathbf{x}})d\bar{\mathbf{x}}=\\
&=\frac{1}{p_{\K}(k)}\int_{\mathcal{R}(\Lambda_3)} \mathds{1}_{\{s=g(\bar{\mathbf{x}}),k=f(\bar{\mathbf{x}})\}} p_{\Z^n|\bar{\X}^n}(\mathbf{z}|\bar{\mathbf{x}})p_{\bar{\X}^n}(\bar{\mathbf{x}})d\bar{\mathbf{x}}=\\
&=\frac{1}{p_{\K}(k)}\int_{\mathcal{V}(k+s)} p_{\Z^n\bar{\X}^n}(\mathbf{z},\bar{\mathbf{x}})d\bar{\mathbf{x}}
\end{align*}
}%
Similarly,
$$p_{\mathsf{S}\Z^n}(s,\mathbf{z})=\sum_{k \in \mathcal{K}_n} \int_{\mathcal{V}(k+s)} p_{\Z^n\bar{\X}^n}(\mathbf{z},\bar{\mathbf{x}})d\bar{\mathbf{x}}.$$
From the bounds in Eq.~(\ref{bound_K}) and Eq.~(\ref{bound_z}), and noticing that $\sigma_2 \leq \sigma_x$ in Eq.~(\ref{variances}), we find
\begin{align*}
& \abs{\int_{\mathcal{V}(k+s)} \frac{p_{\bar{\X}^n\Z^n}(\bar{\mathbf{x}},\mathbf{z})d\bar{\mathbf{x}}}{p_{\K}(k)} -\frac{p_{\Z^n}(\mathbf{z})}{e^{nR_P}}} \leq \frac{4\epsilon_{\Lambda_3}(\sigma_2)}{e^{nR_P}} p_{\Z^n}(\mathbf{z}),\\
& \abs{\sum_{k \in \mathcal{K}_n} \int_{\mathcal{V}(k+s)} p_{\bar{\X}^n\Z^n}(\bar{\mathbf{x}},\mathbf{z})d\bar{\mathbf{x}} -\frac{p_{\Z^n}(\mathbf{z})}{e^{nR_P}}} \leq \frac{\epsilon_{\Lambda_3}(\sigma_2)}{e^{nR_P}} p_{\Z^n}(\mathbf{z})
\end{align*}
provided that $\epsilon_{\Lambda_3}(\sigma_2) \leq \frac{1}{2}$.
Consequently,
$$\mathbb{V}(p_{\mathsf{S}\Z^n|\K=k},p_{\mathsf{S}\Z^n}) \leq \sum_{s} \int_{\R^n} \frac{5 \epsilon_{\Lambda_3}(\sigma_2)}{e^{nR_P}} p_{\Z^n}(\mathbf{z})d\mathbf{z} =5\epsilon_{\Lambda_3}(\sigma_2).$$
Therefore $d_{\av} \leq 5 \epsilon_{\Lambda_3}(\sigma_2)$. If $\Lambda_3$ is secrecy-good, we find 
$$\mathbb{I}(\K;\mathsf{S},\Z^n) \leq 5\epsilon_{\Lambda_3}(\sigma_2)(nR_K-\log 5\epsilon_{\Lambda_3}(\sigma_2)) \to 0.$$
(Actually, $\epsilon_{\Lambda_3}(\sigma_2)=o\left(\frac{1}{n}\right)$ is enough.)

\subsubsection*{Reliability}
Let us analyze the error probability
$\mathbb{P}\{\K \neq \hat{\K}\}$.
Note that
$\K=\hat{\K}$
if
$\hat{\X}^n_Q=\X^n_Q$.
Since
$\X^n_Q=\mathsf{S}+Q_{\Lambda_2}(\X^n_Q),$
we have
\begin{equation*}
\begin{split}
\hat{\X}^n_Q=\X^n_Q &\Longleftrightarrow Q_{\Lambda_2}\left(\rho_{xy}\frac{\sigma_x}{\sigma_y}\Y^n-\mathsf{S}\right)=Q_{\Lambda_2}(\X^n_Q).
\end{split}
\end{equation*}
Since
\begin{equation*}
\begin{split}
Q_{\Lambda_2}\left(\rho_{xy}\frac{\sigma_x}{\sigma_y} \Y^n-\mathsf{S}\right) &= Q_{\Lambda_2}\left(\rho_{xy}\frac{\sigma_x}{\sigma_y} \Y^n-\X^n_Q + Q_{\Lambda_2}(\X^n_Q)\right) \\
&= Q_{\Lambda_2}\left(\rho_{xy}\frac{\sigma_x}{\sigma_y} \Y^n-\X^n_Q\right) + Q_{\Lambda_2}(\X^n_Q)
\end{split}
\end{equation*}
we derive
\begin{equation*}
\begin{split}
\hat{\X}^n_Q=\X^n_Q &\Longleftrightarrow Q_{\Lambda_2}\left(\rho_{xy}\frac{\sigma_x}{\sigma_y} \Y^n -\X^n_Q\right)=0 \\ &\Longleftrightarrow Q_{\Lambda_2}(\mathsf{E}^n_Q -\mathsf{W}_1^n)=0.
\end{split}
\end{equation*}
When
$\epsilon_{\Lambda_1}(\sigma_x)$ and $\epsilon_{\Lambda_1}(\rho_{xy}\sigma_x)$ are small, $\mathsf{E}^n_Q=\X^n \Mod \mathcal{V}(\Lambda_1)$ is almost uniformly distributed on $\mathcal{V}(\Lambda_1)$ and almost independent of $\mathsf{W}_1^n$.
The variance per dimension of $\mathsf{E}^n_Q-\mathsf{W}_1^n$ is asymptotically $G(\Lambda_1)V_1^{2/n}+\sigma_1^2$, where $G(\Lambda_1)$ is the normalized second moment. According to \cite{Zamir02}, if $\Lambda_1$ is good for quantization, then the effect of $\mathsf{E}^n_Q$ on the decoding error probability is subexponential in $n$ relative to the AWGN of the same power.

By the AWGN-goodness of $\Lambda_2$, the error probability
\[
\mathbb{P}\{\K \neq \hat{\K}\} \leq \mathbb{P} \{ Q_{\Lambda_2}(\mathsf{E}^n_Q -\mathsf{W}_1^n)\neq 0\}
\]
will vanish exponentially as long as
\begin{equation*}
\frac{V_2^{2/n}}{\frac{1}{2\pi e}V_1^{2/n}+\sigma_1^2} > 2\pi e.
\end{equation*}

On the other hand, the secrecy-goodness of $\Lambda_3$ requires
\begin{equation} \label{secrecy_goodness_condition}
V_3^{\frac{2}{n}}/\sigma_2^2 < {2\pi}.
\end{equation}

Therefore, the rate of the secret key is bounded by
\begin{equation}\label{key-rate}
R_K < \frac{1}{n}\log\left(\frac{V_3}{V_2}\right) = \frac{1}{2}\log\left(\frac{\sigma_2^2}{\frac{1}{2\pi e}V_1^{2/n}+\sigma_1^2}\right) - \frac{1}{2}.
\end{equation}
This corresponds to the rate of public communication
\begin{equation}\label{public-rate}
R_P = \frac{1}{n}\log\left(\frac{V_2}{V_1}\right) > \frac{1}{2}\log\left(1 + \frac{2\pi e\sigma_1^2}{V_1^{2/n}}\right).
\end{equation}

If we make $\Lambda_1$ sufficiently fine such that $G(\Lambda_1)V_1^{2/n} \ll\sigma_1^2$, then the key rate approaches
\begin{equation} \label{quasi_optimal_rate}
R_K < \frac{1}{2}\log\left(\frac{\sigma_2^2}{\sigma_1^2}\right) - \frac{1}{2}.
\end{equation}
For degraded sources, i.e. $\sigma_1<\sigma_2$ or equivalently $\rho_{xy}>\rho_{xz}$~\cite[Lemma 6]{WaOh10}, the secret key rate is upper-bounded by $\mathbb{I}(\X;\Y)-\mathbb{I}(\X;\Z)$. The key-rate in Eq.~(\ref{quasi_optimal_rate}) is only $1/2$ nat away from this bound.
Achieving this rate requires a high rate $R_P$ of public communication, but $\Lambda_1$ need to be very fine in practice. To see this, we substitute $\frac{1}{2\pi e}V_1^{2/n}=0.1 \sigma_1^2$ in Eq.~(\ref{key-rate}) and~Eq.~\eqref{public-rate}, so that $R_K$ is almost the same as in Eq.~(\ref{quasi_optimal_rate}) while
$R_P > \frac{1}{2}\log\left(1 + 10\right) \approx
1.2$ nats/dimension.
Note, however, that our scheme does not achieve the optimal tradeoff between $R_K$ and $R_P$ identified in~\cite[Theorem 4]{WaOh10} yet. Achieving the optimal tradeoff will be our future work.

\section{Existence of a Sequence of Nested Lattices for Secret Key Generation} \label{existence_section}
We begin by showing the existence of a suitable coarse lattice $\Lambda_3$. For the notions of AWGN-good, Rogers-good and quantization-good lattices we refer the reader to \cite{ErezZamir04, Nazer11}.
Following the same reasoning as in \cite[Appendix III]{LLBS_12}, it can be shown that there exists a sequence $\delta_n \to 0$ and an ensemble of sequences of lattices $\Lambda_3^{(n)}$ with second moment $\sigma^2(\Lambda_3^{(n)})=\frac{\sigma_2^2}{e}$ which are AWGN-good, Rogers-good and quantization-good and such that
\begin{equation} \label{theta_bound}
\mathbb{E}\left[ \Theta_{\Lambda_3^{(n)}}\left(\frac{1}{2\pi \sigma_2^2}\right)\right] \leq 1 + \delta_n + \frac{(2\pi \sigma_2^2)^{\frac{n}{2}}}{V_3^{(n)}}.
\end{equation}
Quantization-goodness then implies that $G(\Lambda_3^{(n)})=\sigma^2(\Lambda_3^{(n)})/(V_3^{(n)})^{\frac{2}{n}} \to 1/2\pi e$, and consequently $V_3^{(n)}$ tends to $2\pi \sigma_2^2$ from below, as required to achieve optimal rate while satisfying the condition in Eq.~(\ref{secrecy_goodness_condition}).
From the average bound in Eq.~(\ref{theta_bound}), and recalling the relation in Eq.~(\ref{flatness_factor_expression}) between theta series and flatness factor, we can deduce the existence of a sequence $\Lambda_3^{(n)}$ which
is also secrecy-good.

By applying twice the technique in \cite[Section VII]{ErezZamir04}, and its extension in \cite{Krithivasan07}, we can find two sequences of fine lattices $\Lambda_1^{(n)}$, $\Lambda_2^{(n)}$ with $\Lambda_1^{(n)} \supset \Lambda_2^{(n)} \supset \Lambda_3^{(n)}$ which are also Rogers, quantization and AWGN-good and such that the volume ratios are arbitrarily close to the bounds in Eq.~(\ref{key-rate}) and Eq.~(\ref{public-rate}). Note that since $\epsilon_{\Lambda'}(\sigma_2) \leq \epsilon_{\Lambda}(\sigma_2)$ whenever $\Lambda \subset \Lambda'$, the lattices  $\Lambda_2^{(n)}$, $\Lambda_1^{(n)}$ are also secrecy-good with respect to $\sigma_2$.

\section*{Acknowledgment}

The work of M. Bloch was supported in part by the CNRS grant PEPS PhySecNet.

\footnotesize
\bibliographystyle{IEEEtran}
\bibliography{IEEEabrv,lingbib}

\end{document}